%% file: paper.tex
\documentclass[twocolumn,showpacs,aps,prl,preprintnumbers,superscriptaddress]{revtex4}

\usepackage{graphicx}
\usepackage{dcolumn}
\usepackage{amsmath}
\usepackage{epsfig}

\long\def\inst#1{\par\nobreak\kern 4pt\nobreak
    {\itshape #1}\par\vskip 10pt plus 3pt minus 3pt}
\RequirePackage{xspace}
\usepackage{relsize}

\def\fL   {\ensuremath{ f_L }}

\def\qqbar {\ensuremath{q\overline q}\xspace}
\def\babar{\mbox{\slshape B\kern-0.1em{\smaller A}\kern-0.1em
    B\kern-0.1em{\smaller A\kern-0.2em R}}}
\def\Dbar    {\kern 0.18em\overline{\kern -0.18em D}{}\xspace}
\def\Bbar    {\kern 0.18em\overline{\kern -0.18em B}{}\xspace}

\def\BB      {\ensuremath{B\Bbar}\xspace} 
\def\Bz      {\ensuremath{B^0}\xspace}
\def\Bzb     {\ensuremath{\Bbar^0}\xspace}
\def\BzBzb   {\ensuremath{\Bz {\kern -0.16em \Bzb}}\xspace}
\def\Bu      {\ensuremath{B^+}\xspace}
\def\Bub     {\ensuremath{B^-}\xspace}

\def\BpBm    {\ensuremath{\Bu {\kern -0.16em \Bub}}\xspace}

\newcommand{\optbar}[1]{\shortstack{{\tiny (\rule[.4ex]{1em}{.1mm})}
  \\ [-.7ex] $#1$}}
\def\BorBbar    {\kern 0.18em\optbar{\kern -0.18em B}{}\xspace}
\def\DorDbar    {\kern 0.18em\optbar{\kern -0.18em D}{}\xspace}
\def\KorKbar    {\kern 0.18em\optbar{\kern -0.18em K}{}\xspace}
\def\CP                {\ensuremath{C\!P}\xspace}
\def\pep2{PEP-II}
\mathchardef\Upsilon="7107
\def\Y#1S{\ensuremath{\Upsilon{(#1S)}}\xspace}% no space before {...}!
\def\FourS {\Y4S}
\def\pip   {\ensuremath{\pi^+}\xspace}
\def\pim   {\ensuremath{\pi^-}\xspace}

\def\pipm  {\ensuremath{\pi^\pm}\xspace}
\def\pimp  {\ensuremath{\pi^\mp}\xspace}

\def\invfb   {\ensuremath{\mbox{\,fb}^{-1}}\xspace}

\def\B       {\ensuremath{B}\xspace}
\def\mes        {\mbox{$m_{\rm ES}$}\xspace}
\def\DeltaE     {\mbox{$\Delta E$}\xspace}

\def\Kp    {\ensuremath{K^+}\xspace}
\def\Km    {\ensuremath{K^-}\xspace}

\def\Dstar   {\ensuremath{D^*}\xspace}
\def\Dmeson   {\ensuremath{D}\xspace}
\def\Dss     {\ensuremath{D^{*\pm}_s}\xspace}

\newcommand{\jprlBase}       {Phys.\ Rev.\ Lett.\xspace}
\newcommand{\jprl}      [1]  {\jprlBase\ {\bf #1}}
\newcommand{\jplBase}        {Phys.\ Lett.\xspace}
\newcommand{\jpl}       [1]  {\jplBase\ {\bf #1}}
\newcommand{\plb}       [1]  {\jplBase\ B~{\bf #1}}
\newcommand{\jprBase}        {Phys.\ Rev.\xspace}
\newcommand{\jprd}      [1]  {\jprBase\ D~{\bf #1}}
\newcommand{\npBase}         {Nucl.\ Phys.\xspace}
\newcommand{\npb}       [1]  {\npBase\ B~{\bf #1}}
\newcommand{\nimBaseA}       {Nucl.\ Instrum.\ Methods Phys.\ Res., Sect.\ A\xspace}
\newcommand{\nima}      [1]  {\nimBaseA~{\bf #1}}
\newcommand{\progtp}    [1]  {{Prog.\ Theor.\ Phys.\ {\bf #1}}}

\def\cm   {\ensuremath{{\rm \,cm}}\xspace}

\def\piz    {\ensuremath{\pi^0}\xspace}

\def\mkpi   {\ensuremath{m_{K\pi}}}

\def\VV     {\ensuremath{VV}\xspace} 

\def\calB    {\ensuremath{{\cal B}}\xspace}
\def\calF    {\ensuremath{{\cal F}}\xspace}

\def\Kbar    {\kern 0.2em\overline{\kern -0.2em K}{}\xspace}

\def\Kz      {\ensuremath{K^0}\xspace}

\def\Kstar   {\ensuremath{K^{*}}\xspace}
\def\Kstarp  {\ensuremath{K^{*+}}\xspace}
\def\Kstarm  {\ensuremath{K^{*-}}\xspace}
\def\Kstarpm {\ensuremath{K^{*\pm}}\xspace}
\def\Kstarmp {\ensuremath{K^{*\mp}}\xspace}
\def\Kstarz  {\ensuremath{K^{*0}}\xspace}
\def\Kstarzb {\ensuremath{\Kbar^{*0}}\xspace}

\def\KS    {\ensuremath{K^0_{\scriptscriptstyle S}}\xspace} 
\def\Kp    {\ensuremath{K^+}\xspace}
\def\Km    {\ensuremath{K^-}\xspace}
\def\Kpm   {\ensuremath{K^\pm}\xspace}
\def\Kmp   {\ensuremath{K^\mp}\xspace}

\def\btoKstarpKstarm {\ensuremath{\Bz \rightarrow \Kstarp\Kstarm}}

\def\btoKstarpKstarmKz {\ensuremath{\Bz \to \Kstarpm(\to \KS\pipm)\Kstarmp(\to \KS\pimp)}}

\def\btoKstarpKstarmKp {\ensuremath{\Bz \to \Kstarpm(\to \KS\pipm)\Kstarmp(\to \Kmp\piz)}}

\newcommand{\gev}{\ensuremath{\mathrm{\,Ge\kern -0.1em V}}\xspace}
\newcommand{\mev}{\ensuremath{\mathrm{\,Me\kern -0.1em V}}\xspace}
\newcommand{\kev}{\ensuremath{\mathrm{\,ke\kern -0.1em V}}\xspace}
\newcommand{\ev}{\ensuremath{\mathrm{\,e\kern -0.1em V}}\xspace}
\newcommand{\gevc}{\ensuremath{{\mathrm{\,Ge\kern -0.1em V\!/}c}}\xspace}
\newcommand{\mevc}{\ensuremath{{\mathrm{\,Me\kern -0.1em V\!/}c}}\xspace}
\newcommand{\gevcc}{\ensuremath{{\mathrm{\,Ge\kern -0.1em V\!/}c^2}}\xspace}
\newcommand{\mevcc}{\ensuremath{{\mathrm{\,Me\kern -0.1em V\!/}c^2}}\xspace}

\def\etal{{\em et al.}}
\def\epem       {\ensuremath{e^+e^-}\xspace}

\def\cossq    {\ensuremath{\cos^2\theta}}
\def\sinsq    {\ensuremath{\sin^2\theta}}

\def\Bmeson  {\B\ meson}
\def\Bmesons {\B\ mesons}

\def\Bback   {\BB\ background}
\def\Bbacks  {\BB\ backgrounds}

\include{numbers}
\begin{document}

\preprint{\babar-PUB-08/019}
\preprint{SLAC-PUB-13275}
\preprint{arXiv:0806.4467}
%\preprint{\babar\ Analysis Document 1959 Version 8}

\title{
\large \bfseries \boldmath Search for \btoKstarpKstarm }

\date{\today}
\input{authors_may2008}

\begin{abstract}
We report the results of a search for the decay \btoKstarpKstarm\ 
with a sample of \nbb\ million \BB\ pairs
collected with the \babar\ detector at the PEP-II asymmetric-energy
\epem\ collider at the Stanford Linear Accelerator Center. We
 obtain an upper limit at the 90\% confidence level on the branching
fraction for $\calB\ (\btoKstarpKstarm) < \fkcup\times 10^{-6}$,
assuming the decay is fully longitudinally polarized.
\end{abstract}

% PACS, the Physics and Astronomy Classification Scheme.

\maketitle

%%%%%%%%%%%%%%%%%%%%%%%%%%%%%%%%%%%%%%%%%%%%%%%%%%%%%%%%%%%%%%%
% THEORY
%%%%%%%%%%%%%%%%%%%%%%%%%%%%%%%%%%%%%%%%%%%%%%%%%%%%%%%%%%%%%%%

The study of the branching fractions and angular distributions of
\Bmeson\ decays to hadronic final states without a charm quark probes
the dynamics of both weak and strong interactions, and plays an
important role in understanding \CP\ violation.  Improved experimental
measurements of these charmless decays, combined with theoretical
developments, can provide significant constraints on the
Cabibbo-Kobayashi-Maskawa (CKM) matrix parameters~\cite{bib:ckm} and
uncover evidence for physics beyond the standard
model~\cite{bib:Beneke06,cheng08}.

QCD factorization models predict the angular distribution of the decay
of the \Bmeson\ to two vector particles (VV), as measured by the
longitudinal polarization fraction \fL, to be $\sim 0.9$ for both
tree- and penguin-dominated decays~\cite{bib:prediction}. Two
measurements of the pure penguin \VV\ decay $B\rightarrow \phi K^*$
give \fL $= 0.52\pm0.08\pm0.03$ and \fL $=
0.49\pm0.05\pm0.03$~\cite{bib:phiKst}, while \fL $=
0.81^{+0.10}_{-0.12}\pm0.06$ has recently been measured for the decay
$B^0\rightarrow \Kstarz\Kstarzb$~\cite{bib:KstKst}.  Several attempts
to understand the values of \fL\ within or beyond the standard model
have been made~\cite{bib:theory1}.  Further information about decays
related by $SU(3)$ symmetry may provide insights into this
polarization puzzle and test factorization models.

\begin{figure}[!ht]
\begin{center}
\begin{tabular}{c}
    \epsfig{file=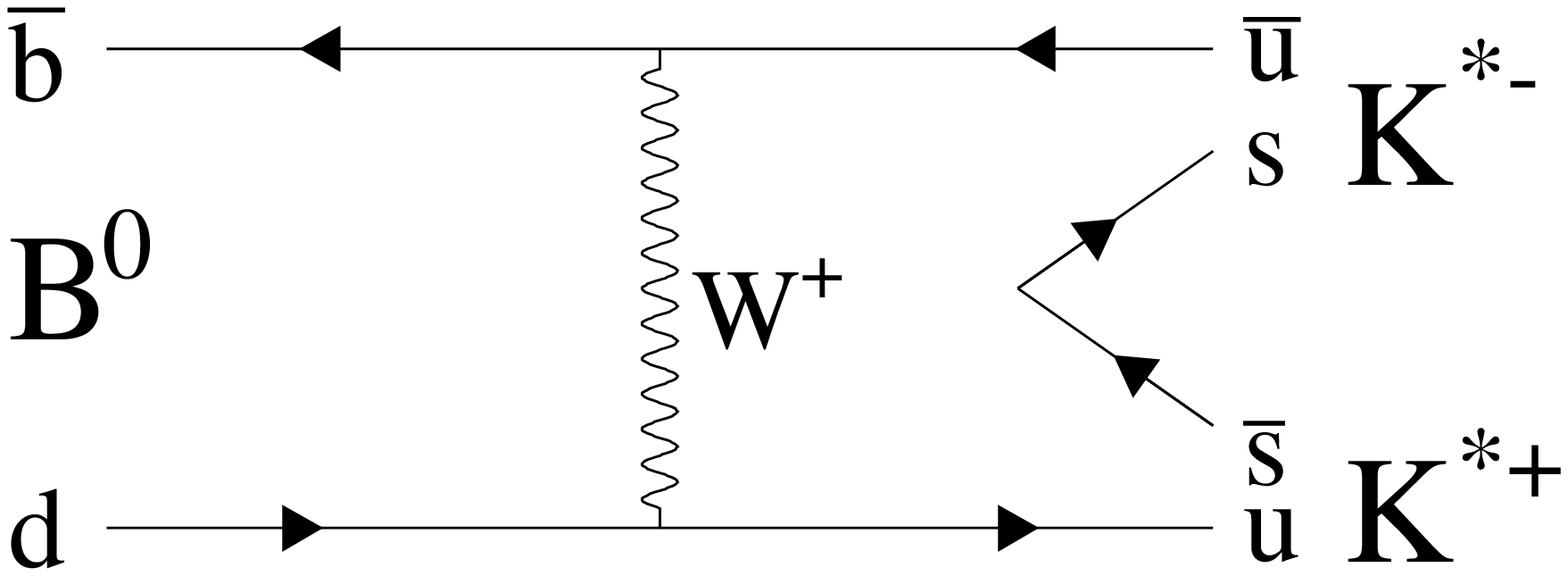,width=0.75\columnwidth}
\end{tabular}
\caption
{The $b \to u$ $W$-exchange diagram for \btoKstarpKstarm.}
\label{fig:feynman}
\end{center}
\end{figure}

The decay \btoKstarpKstarm\ is expected to occur through a
$b\rightarrow u$ quark transition via $W$-exchange, as shown in
Figure~\ref{fig:feynman}, or from final-state interactions. Its
branching fraction is expected to be small, with Beneke, Rohrer and
Yang~\cite{bib:Beneke06} predicting $(0.09^{+0.05+0.12}_{-0.03-0.10})
\times 10^{-6}$, while Cheng and Yang~\cite{cheng08} quote $(0.1 \pm
0.0 \pm 0.1) \times 10^{-6}$, both based on QCD factorization. The
current experimental upper limit on the branching fraction at the 90\%
confidence level (C.L.) is $141 (89) \times
10^{-6}$~\cite{bib:prevcleo}, assuming a fully longitudinally
(transversely) polarized system.  Searches for the related decay $\Bz
\to \Kp\Km$ have produced upper limits on the branching fraction at
the 90\% C.L. in the range $(0.4-0.8)\times 10^{-6}$~\cite{bib:kpkm}.

%%%%%%%%%%%%%%%%%%%%%%%%%%%%%%%%%%%%%%%%%%%%%%%%%%%%%%%%%%%%%
% INTRODUCTION
%%%%%%%%%%%%%%%%%%%%%%%%%%%%%%%%%%%%%%%%%%%%%%%%%%%%%%%%%%%%%%%

%We report measurements of the branching fraction and the longitudinal
% polarization fraction for the decay mode \btoKstarpKstarm, where

We report on a search for the decay mode \btoKstarpKstarm, where
 \Kstarpm\ refers to the $K^{*\pm}(892)$ resonance, without explicit
 consideration of interference from higher mass \Kstar\ states, and place
 an upper limit on the branching fraction. Charge-conjugate modes are
 implied throughout and we assume equal production rates of \BpBm and
 \BzBzb.

%%%%%%%%%%%%%%%%%%%%%%%%%%%%%%%%%%%%%%%%%%%%%%%%%%%%%%%%%%%%%%%%
% DETECTOR AND DATASET
%%%%%%%%%%%%%%%%%%%%%%%%%%%%%%%%%%%%%%%%%%%%%%%%%%%%%%%%%%%%%%%%

This analysis is based on a data sample of \nbb\ million \BB\ pairs,
corresponding to an integrated luminosity of \onreslumi, collected
with the \babar\ detector at the PEP-II
asymmetric-energy \epem\ collider operated at the Stanford Linear
Accelerator Center. The \epem\ center-of-mass
(c.m.) energy is $\sqrt{s} = 10.58$\gev, corresponding to the
\FourS\-resonance mass (on-resonance data). In addition, \offreslumi\
of data collected at 40~\mev\ below the \FourS\-resonance (off-resonance
data) are used for background studies.

%%%%%%%%%%%%%%%%%%%%%%%%%%%%%%%%%%%%%%%%%%%%%%%%%%%%%%%%%%%%%%%%%%
% Detector
%%%%%%%%%%%%%%%%%%%%%%%%%%%%%%%%%%%%%%%%%%%%%%%%%%%%%%%%%%%%%%%%%%

The \babar\ detector is described in detail in Ref.~\cite{bib:babar}.
Charged particles are reconstructed as tracks with a 5-layer silicon
vertex detector (SVT) and a 40-layer drift chamber inside a $1.5$-T
solenoidal magnet. An electromagnetic calorimeter (EMC) comprising
6580 CsI(Tl) crystals is used to identify electrons and photons.  A
ring-imaging Cherenkov detector (DIRC) is used to identify charged
hadrons and to provide additional electron identification
information. The average $K$-$\pi$ separation in the DIRC varies from
12$\sigma$ at a laboratory momentum of 1.5\gevc\ to 2.5$\sigma$ at
4.5\gevc.  Muons are identified by an instrumented magnetic-flux
return (IFR).

%%%%%%%%%%%%%%%%%%%%%%%%%%%%%%%%%%%%%%%%%%%%%%%%%%%%%%%%%%%%%%%%%%
% MC
%%%%%%%%%%%%%%%%%%%%%%%%%%%%%%%%%%%%%%%%%%%%%%%%%%%%%%%%%%%%%%%%%%

%%%%%%%%%%%%%%%%%%%%%%%%%%%%%%%%%%%%%%%%%%%%%%%%%%%%%%%%%%%%%%%%%%
% ANALYSIS METHOD
%%%%%%%%%%%%%%%%%%%%%%%%%%%%%%%%%%%%%%%%%%%%%%%%%%%%%%%%%%%%%%%%%%

The \btoKstarpKstarm\ candidates are reconstructed through the decay
of both $\Kstarpm$ to $\KS\pipm$ or with one $\Kstarpm$ decaying to 
$\KS\pipm$ and
the other to $\Kmp\piz$.  The differential decay rate, after
integrating over the angle between the decay planes of the vector
mesons, for which the acceptance is uniform, is
\begin{eqnarray}
\lefteqn{\frac{1}{\Gamma}\frac{d^2\Gamma}{d\cos\theta_{1}d\cos\theta_{2}} 
\propto} \nonumber \\  &  & \frac{1-f_L}{4}\sin^2\theta_{1}\sin^2\theta_{2} + 
   f_L \cos^2\theta_{1}\cos^2\theta_{2} ,
\label{eq:helicity}
\end{eqnarray}

\noindent where $\theta_{1}$ and $\theta_{2}$ are the helicity angles
of the \Kstarp\ and \Kstarm, defined as the angle between the daughter
kaon (\KS\ or \Kpm) momentum and the direction opposite to the
\Bmeson\ in the \Kstarpm rest frame~\cite{bib:polarization}.

%%%%%%%%%%%%%%%%%%%%%%%%%%%%%%%%%%%%%%%%%%%%%%%%%%%%%%%%%%%%%%%%%%
% EVENT SELECTION
%%%%%%%%%%%%%%%%%%%%%%%%%%%%%%%%%%%%%%%%%%%%%%%%%%%%%%%%%%%%%%%%%%

The charged particles from the \Kstarpm\ decays are required to have
at least 12 hits in the drift chamber and a transverse momentum
greater than 0.1\gevc. The particles are identified as either charged
pions or kaons by measurement of the energy loss in the tracking
devices, the number of photons recorded by the DIRC and
the corresponding Cherenkov angle. These measurements are combined
with additional information from the EMC and IFR detectors, where
appropriate, to reject electrons, muons, and protons.

The \KS\ is reconstructed through its decay to \pip\pim. The \KS\
candidates are required to have a reconstructed mass within 0.01\gevcc\
of the nominal \KS\ mass~\cite{bib:PDG}, a decay vertex separated from
the \Bmeson\ decay vertex by at least twenty times the uncertainty in
the measurement of the vertex position, a flight distance in the
transverse direction of at least 0.3\cm, and the cosine of the angle
between the line joining the $B$ and \KS\ decay vertices and the \KS\
momentum greater than 0.999.

We reconstruct the \piz\ through the decay $\piz\to\gamma\gamma$.
In the laboratory frame, the energy of each photon from the 
\piz\ candidate must be greater than 0.04\gev, the
energy of the \piz\ must be greater than 0.25\gev, and the
reconstructed \piz\ invariant mass is required to be $0.12\le
m_{\gamma\gamma} \le 0.15$\gevcc.

We require the invariant mass of the \Kstarp\ candidates to be $0.792
< \mkpi < 0.992$\gevcc. A \Bmeson\ candidate is formed from the
\Kstarpm\ candidates, with the condition that the \Kstarpm\
candidates originate from the interaction region.

\Bmeson\ candidates are characterized kinematically by the energy
 difference $\DeltaE = E^*_B - \sqrt{s}/2$ and the beam energy-substituted
 mass $m_{\rm ES} = \left [ (s/2+{\bf p}_i\cdot{\bf p}_B)^2/E_i^2-{\bf
 p}_B^2\right ] ^{1/2}$, where $(E_i,{\bf p}_i)$ and $(E_B,{\bf p}_B)$
 are the four-momenta of the \FourS and \Bmeson\ candidate,
 respectively, and the asterisk denotes the \FourS\ rest frame.  For a
 final state with a \piz, the total event sample is taken from the
 region $-0.1 \le \DeltaE \le 0.2$\gev and $5.25 \le \mes \le
 5.29$\gevcc; with no \piz, the signal \DeltaE\ has a smaller width
 and the region $-0.08 \le \DeltaE \le 0.15$\gev is used. The
 asymmetric \DeltaE\ criteria are applied to remove backgrounds from
 charm decays which occur in the negative \DeltaE\ region. In both
 cases, events outside the region $|\DeltaE|\le 0.07\gev$ and $5.27
 \le \mes \le 5.29\gevcc$ are used to characterize the background.

We suppress the background from decays to charmed states by forming
the invariant mass, $m_{\Dmeson}$, from combinations of three out of
the four daughter particles' four-momenta. The event is rejected if $1.845 <
m_{\Dmeson} < 1.895$\gevcc\ and the charge and particle type of the
tracks are consistent with a decay from a \Dmeson\ meson.  We reduce
backgrounds from $\Bz\to \phi\Kstarz$ by assigning the kaon mass to
the pion candidate and rejecting the event if the combined invariant
mass of the two charged tracks is between $1.00$ and $1.04$\gevcc.
Finally, to reduce the continuum background and avoid the region where
the reconstruction efficiency falls off rapidly for low momentum
tracks, we require the cosine of the helicity angle of the \Kstarpm\
candidates to be in the range $-1.0\le \cos(\theta) \le 0.9$ for
states without a \piz\ and $-0.9\le \cos(\theta) \le 0.9$ for decays
with a \piz.

To reject the dominant background consisting of light-quark \qqbar
$(q = u,d,s,c)$ continuum events, we require $|\cos\theta_T|<0.8$,
where $\theta_T$ is the angle, in the c.m.\ frame, between the thrust
axis~\cite{bib:thrust} of the \Bmeson\ and that formed from the other
tracks and neutral clusters in the event. Signal events have a flat
distribution in $|\cos\theta_T|$, while continuum events peak at
$1$.

%After the application of the selection criteria, the average number of
%Monte Carlo (MC) signal candidates per selected data event is 1.08
%(1.02) for fully longitudinally (transversely) polarized decays with
%no \piz\ in the final state and 1.18 (1.10) for decays with one \piz\
%in the final state. 

We use Monte Carlo (MC) simulations of the signal decay to estimate
the number of signal candidates per event.  After the application of
the selection criteria, the average number of signal candidates per
event is predicted to be 1.08 (1.02) for fully longitudinally
(transversely) polarized decays with no \piz\ in the final state and
1.18 (1.10) for decays with one \piz\ in the final state.  A single
candidate per event is chosen as the one whose fitted decay vertex has
the smallest $\chi^2$.  MC simulations also show that up to 7\%
(2.4\%) of longitudinally (transversely) polarized signal events with
no \piz\ are misreconstructed, with one or more tracks originating
from the other \Bmeson\ in the event.  In the case of signal events
with one \piz, the number of misreconstructed candidates is 11\%
(4.3\%) for longitudinally (transversely) polarized signal events.

 We create a Fisher discriminant \calF\ to be used in the
maximum-likelihood (ML) fit, constructed from a linear combination of
five variables: the polar angles of the \Bmeson\ momentum vector and
the \Bmeson\ thrust axis with respect to the beam axis, the ratio of
the second- and zeroth-order momentum-weighted Legendre polynomial
moments of the energy flow around the \Bmeson\ thrust axis in the
c.m.\ frame~\cite{bib:Legendre}, the flavor of the other \Bmeson\ as
reported by a multivariate tagging algorithm~\cite{bib:tagging}, and
the boost-corrected proper-time difference between the decays of the
two \Bmesons\ divided by its variance. The second \Bmeson\ is formed
by creating a vertex from the remaining tracks that are consistent
with originating from the interaction region. The Fisher discriminant 
is trained using MC for signal and \qqbar\ continuum MC, off-resonance
data and data
outside the signal region for the background.

%%%%%%%%%%%%%%%%%%%%%%%%%%%%%%%%%%%%%%%%%%%%%%%%%%%%%%%%%%%
% Likelihood Fit
%%%%%%%%%%%%%%%%%%%%%%%%%%%%%%%%%%%%%%%%%%%%%%%%%%%%%%%%%%%

We use an extended unbinned ML fit to extract
the signal yield and polarization
simultaneously for each mode. The extended likelihood function is
\begin{equation}
{\mathcal L} = \frac{1}{N!}\exp{\left(-\sum_{j}n_{j}\right)}
\prod_{i=1}^N\left[\sum_{j}n_{j}{\mathcal
    P}_{j}(\vec{x}_i;\vec{\alpha}_j)\right]\!.
\end{equation}

\noindent We define the likelihood ${\cal L}_i$ for each event
candidate $i$ as the sum of $n_j {\cal P}_j(\vec x_i; \vec \alpha_j)$
over three hypotheses $j$ (signal, \qqbar\ background and
\Bbacks\ as discussed below), where ${\cal P}_j(\vec x_i; \vec
\alpha_j)$ is the product of the probability density functions (PDFs)
for hypothesis $j$ evaluated for the $i$-th event's measured variables
$\vec x_i$, $n_j$ is the yield for hypothesis $j$, and $N$ is the
total number of events in the sample. The quantities $\vec \alpha_j$
represent parameters in the expected distributions of the measured
variables for each hypothesis $j$.  Each discriminating variable $\vec
x_i$ in the likelihood function is modeled with a PDF, where the
parameters $\vec \alpha_j$ are extracted from MC simulation,
off-resonance data, or (\mes, \DeltaE) sideband data.

The seven variables $\vec x_i$ used in the fit are \mes, \DeltaE,
\calF, and the invariant masses and cosines of the helicity angle of
the two \Kstarpm\ candidates. Since the correlations among the fitted
input variables are found to be on average $\sim 1\%$, with a maximum
of 5\%, we take each ${\cal P}_j$ to be the product of the PDFs for
the separate variables. The effect of neglecting correlations is
evaluated by fitting ensembles of simulated experiments in which we
embed signal and background events randomly extracted from
fully-simulated MC samples. Any observed fit bias is then subtracted from
the fitted yield.

For the final state with no \piz, the two invariant mass and helicity
angle distributions for each \Kstarpm\ meson are indistinguishable and
so we use the same PDF parameters for both \Kstarpm\ candidates; for
the final state with a \piz, we use separate PDFs for
$\Kstarpm\to\KS\pipm$ and $\Kstarmp\to\Kmp\piz$.  For the signal, we
use a relativistic Breit-Wigner for the \Kstarpm\ invariant mass and a
sum of two Gaussians for \mes\ and \DeltaE.  The longitudinal
(transverse) helicity angle distributions are described with a \cossq\
(\sinsq) function corrected for changes in efficiency as a function of
helicity angle. The correction also accounts for the reduction in
efficiency at a helicity of $\sim 0.78$ introduced indirectly by the
criteria used to veto \Dmeson\ mesons. The \Bbacks\ use an empirical
non-parametric function for \DeltaE, the masses and helicity
angles. The continuum and the \Bback\ \mes\ shapes are described by the
function $x\sqrt{1-x^2}\exp[-\xi (1-x^2)]$ (with $x=\mes/E^*_B$ and
$\xi$ a free parameter)~\cite{bib:argus} and a first- or third-order
polynomial is used for \DeltaE\ and the helicity angles,
respectively. The continuum invariant mass distributions contain real
\Kstarpm\ candidates; we model the peaking mass component using the
parameters extracted from the fit to the signal invariant mass
distributions together with a second-order polynomial to represent the
non-peaking component.  The Fisher distributions are modeled using an
asymmetric Gaussian for all hypotheses.

 \Bbacks\ that remain after the event selection criteria have been
 applied are identified and modeled using MC simulation based
 on the full physics and detector models~\cite{bib:geant}. There are
 no significant charmless \Bbacks. The charm \Bbacks\ are effectively
 suppressed by applying the veto on \Dmeson\ meson mass described
 above. The remaining charm \Bback\ events are mostly single
 candidates formed from the decay products of a \Dmeson, \Dstar\ or
 \Dss, together with another track from the event. Given the
 uncertainty in the polarization and branching fractions of these
 backgrounds, we allow the \Bback\ yield to float in the fit.

The continuum background PDF parameters that are allowed to vary are
the \calF\ peak position, $\xi$ for \mes, the slope of \DeltaE, and
the polynomial coefficients and normalizations describing the mass and
helicity angle distributions. We fit for the branching fraction \calB\ 
and \fL\ directly and exploit the fact that \calB\ is less
correlated with \fL\ than is either the yield or efficiency taken
separately.  We validate the fitting procedure and extract fitting
biases by applying the fit to ensembles of simulated experiments using
the extracted fitted yields from data. The \qqbar\ component is drawn
from the PDF, and the signal and \Bback\ events are randomly sampled
from the fully simulated MC samples.

%%%%%%%%%%%%%%%%%%%%%%%%%%%%%%%%%%%%%%%%%%%%%%%%%%%%%%%%%%%%%%%
% PHYSICS RESULTS
%%%%%%%%%%%%%%%%%%%%%%%%%%%%%%%%%%%%%%%%%%%%%%%%%%%%%%%%%%%%%%%

The total event sample consists of \kzntot\ and \kpntot\ events for
\btoKstarpKstarm\ with zero or one \piz\ in the final state,
respectively. The corresponding signal event yield is \kznsig\ and
\kpnsig\ and the longitudinal polarization \fL\ is \kzfl\ and \kpfl,
respectively. Given the large errors on \fL, we repeat the analysis
with \fL\ set to 1.0; this gives the most conservative 90\% confidence
level upper limit on the branching fractions.  The results of the ML
fits with $\fL=1.0$ are summarized in Table~\ref{tab:results}. The
\Bback\ yield agrees with the MC prediction within the large
statistical errors.  We compute the branching fractions \calB\ by
subtracting the ML fit bias from the fitted yield and dividing the
result by the number of \BB\ pairs and by the reconstruction
efficiency, $\epsilon$, times $\calB(\Kz\to\KS\to\pip\pim) = 0.5\times
(69.20\pm0.05)\%$ and $\calB(\piz\to\gamma\gamma) = (98.80\pm0.03)\%$.
The significance $S$ of the signal is defined as $S=2\Delta\ln {\cal
L}$, where $\Delta\ln {\cal L}$ is the change in likelihood from the
maximum value when the number of signal events is set to zero,
corrected for the systematic errors defined below.

The significance of the \btoKstarpKstarm\ branching fraction is
$\fkcsig \sigma$, including statistical and systematic uncertainties.
The 90\% C.L. branching fraction upper limit (${\cal B}_{\rm UL}$) is
determined by combining the likelihoods from the two fits and
integrating the total likelihood distribution (taking into account
correlated and uncorrelated systematic uncertainties) as a function
of the branching fraction from 0 to ${\cal B}_{\rm UL}$, so that
$\int^{{\cal B}_{\rm UL}}_0 {\cal L}d{\cal B} = 0.9 \int^\infty_0
{\cal L}d{\cal B}$.

\begin{figure}[!ht]
\centerline{
\setlength{\epsfxsize}{0.5\linewidth}\leavevmode\epsfbox{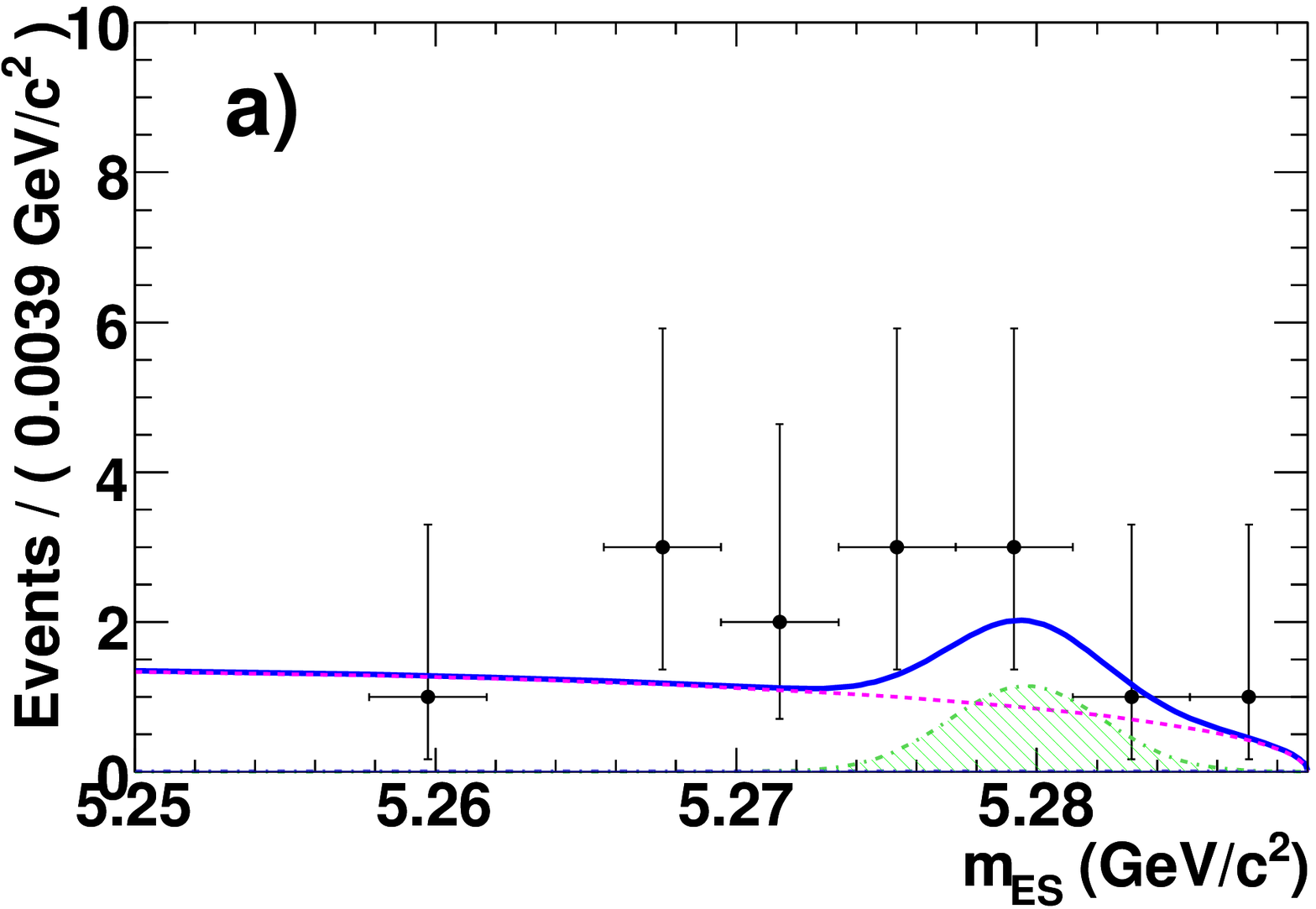}
\setlength{\epsfxsize}{0.5\linewidth}\leavevmode\epsfbox{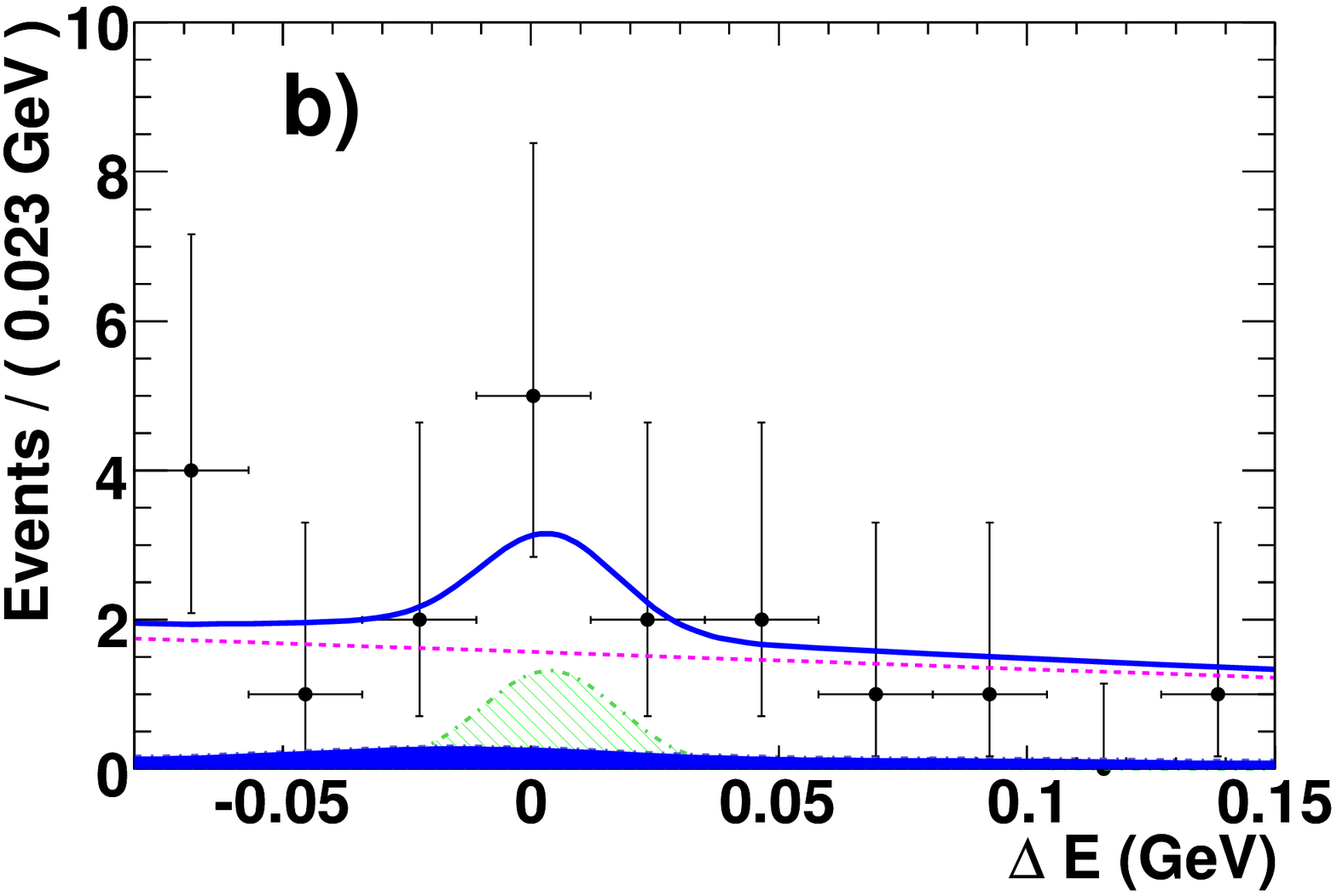}
}
\centerline{
\setlength{\epsfxsize}{0.5\linewidth}\leavevmode\epsfbox{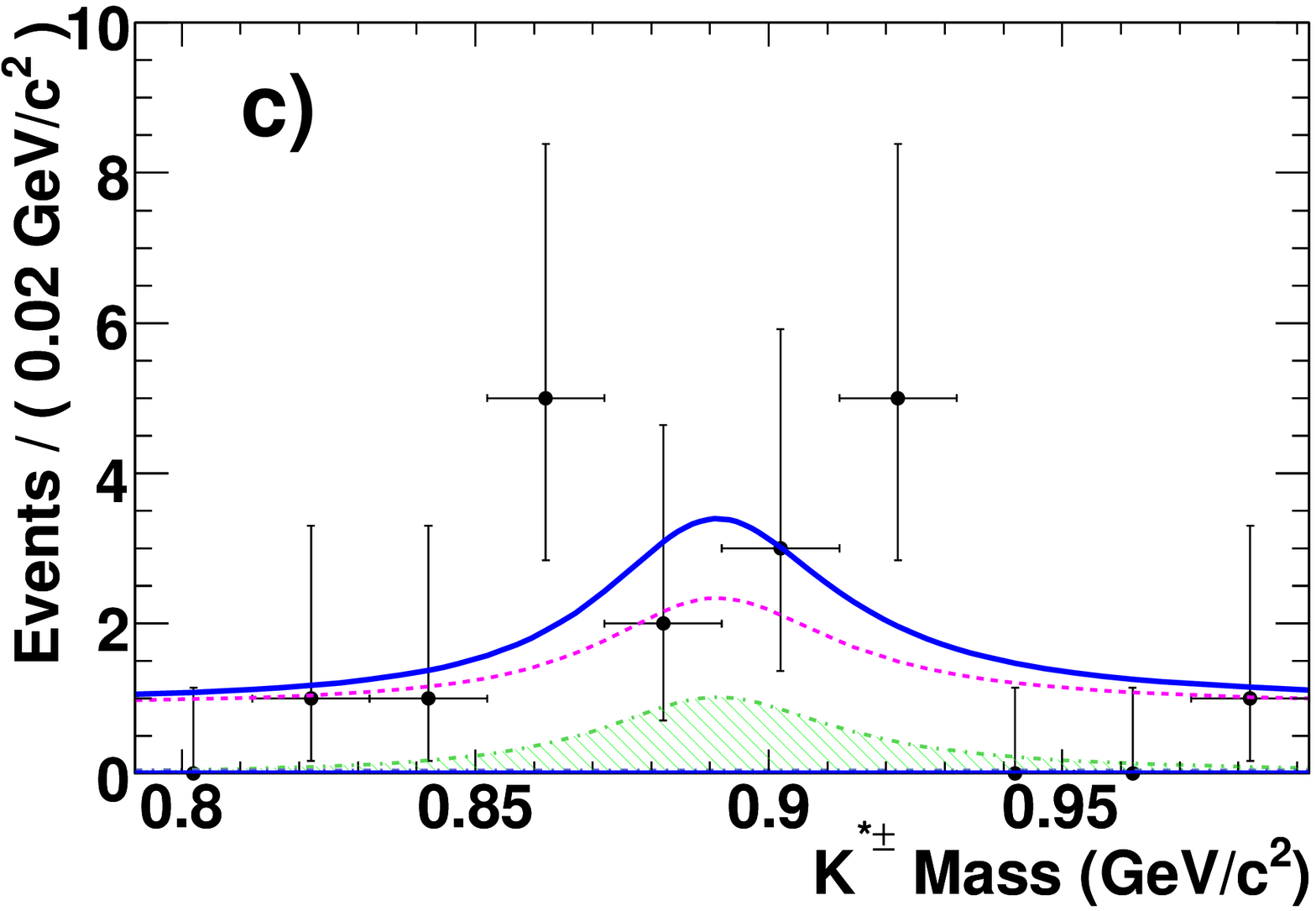}
\setlength{\epsfxsize}{0.5\linewidth}\leavevmode\epsfbox{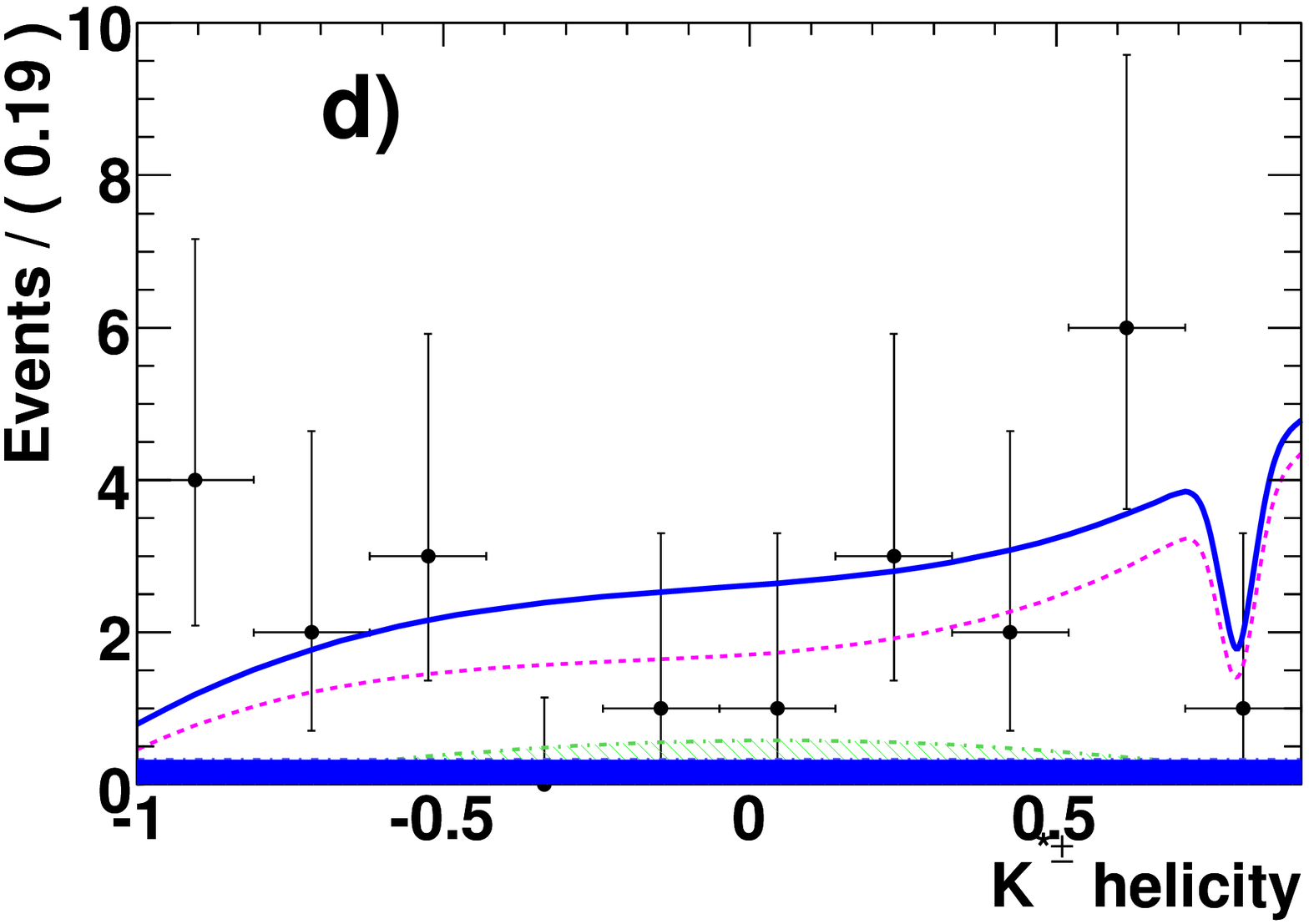}
}
\vspace{-0.3cm}
\caption{\label{fig:proj1} Projections of the
multidimensional fit onto (a) \mes; (b) \DeltaE; (c) \Kstarpm\ mass;
and (d) cosine of \Kstarpm\ helicity angle for \btoKstarpKstarmKz\ 
events selected with a requirement on the signal-to-total likelihood
probability ratio, optimized for each variable, with the plotted
variable excluded.  The points with error bars show the data; the
solid line shows signal-plus-background; the
dashed line is the continuum background; the hatched region is the
signal; and the shaded region is the \Bback.}
\label{fig:fig01}
\end{figure}

\begin{figure}[t]
\centerline{
\setlength{\epsfxsize}{0.5\linewidth}\leavevmode\epsfbox{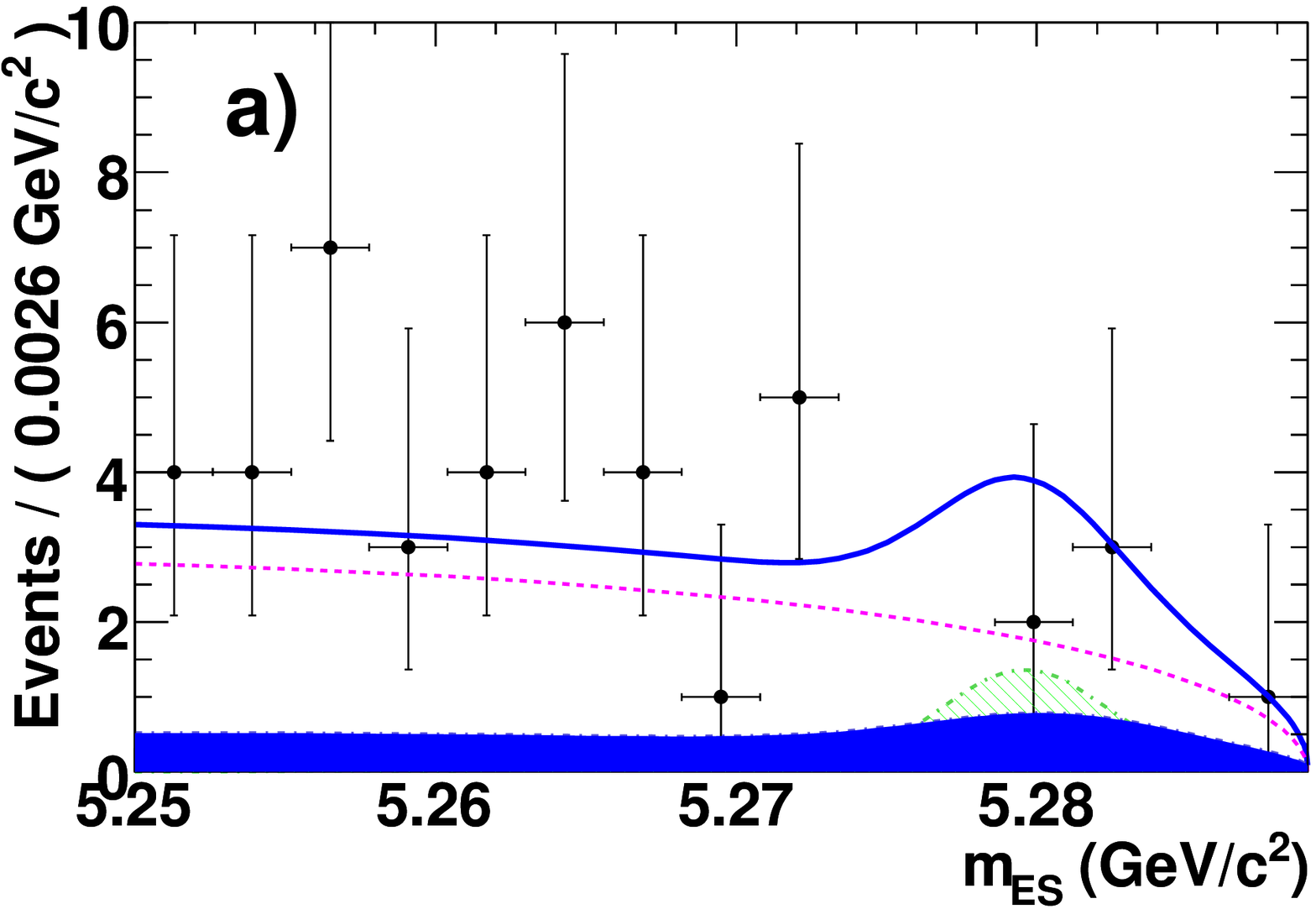}
\setlength{\epsfxsize}{0.5\linewidth}\leavevmode\epsfbox{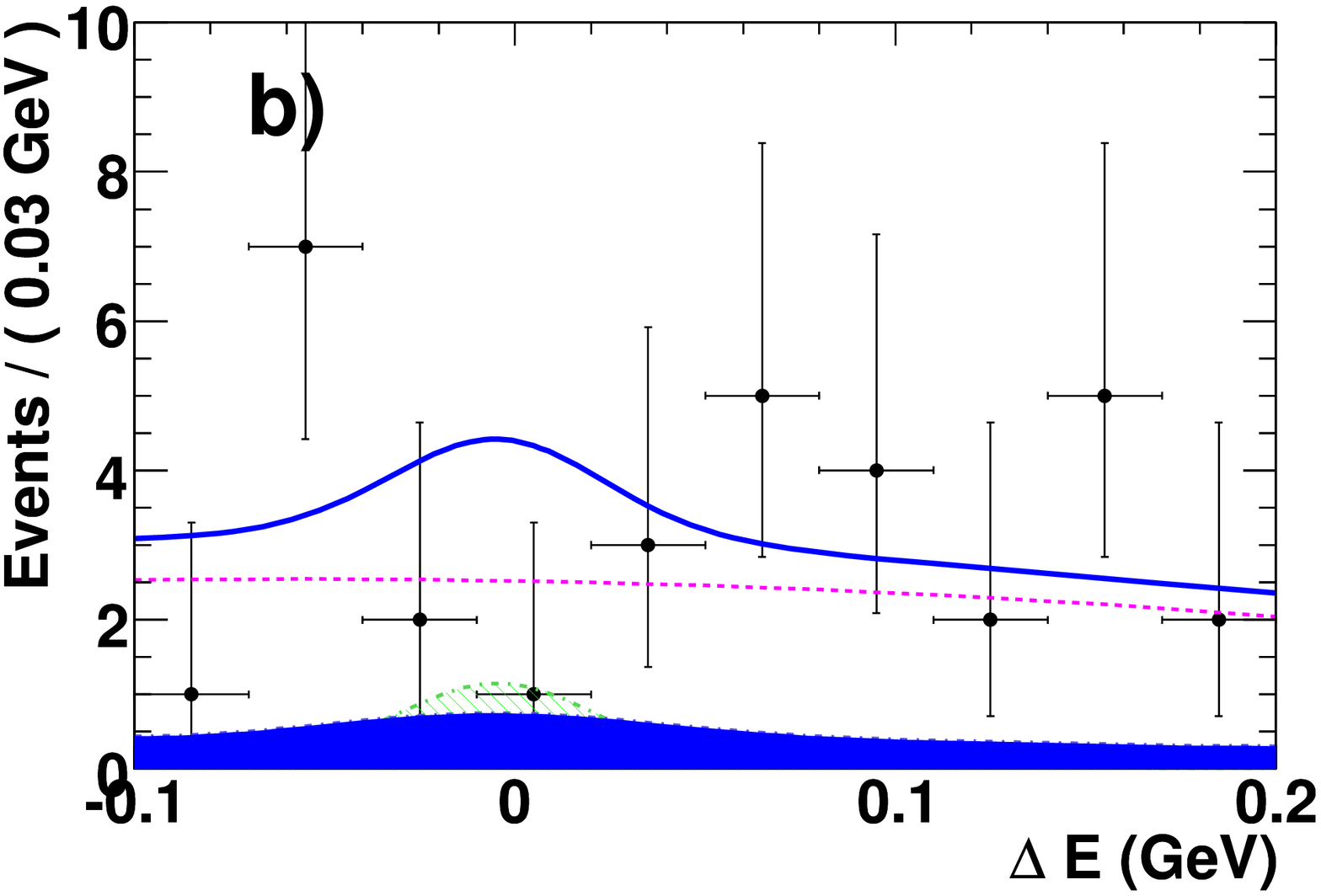}
}
\centerline{
\setlength{\epsfxsize}{0.5\linewidth}\leavevmode\epsfbox{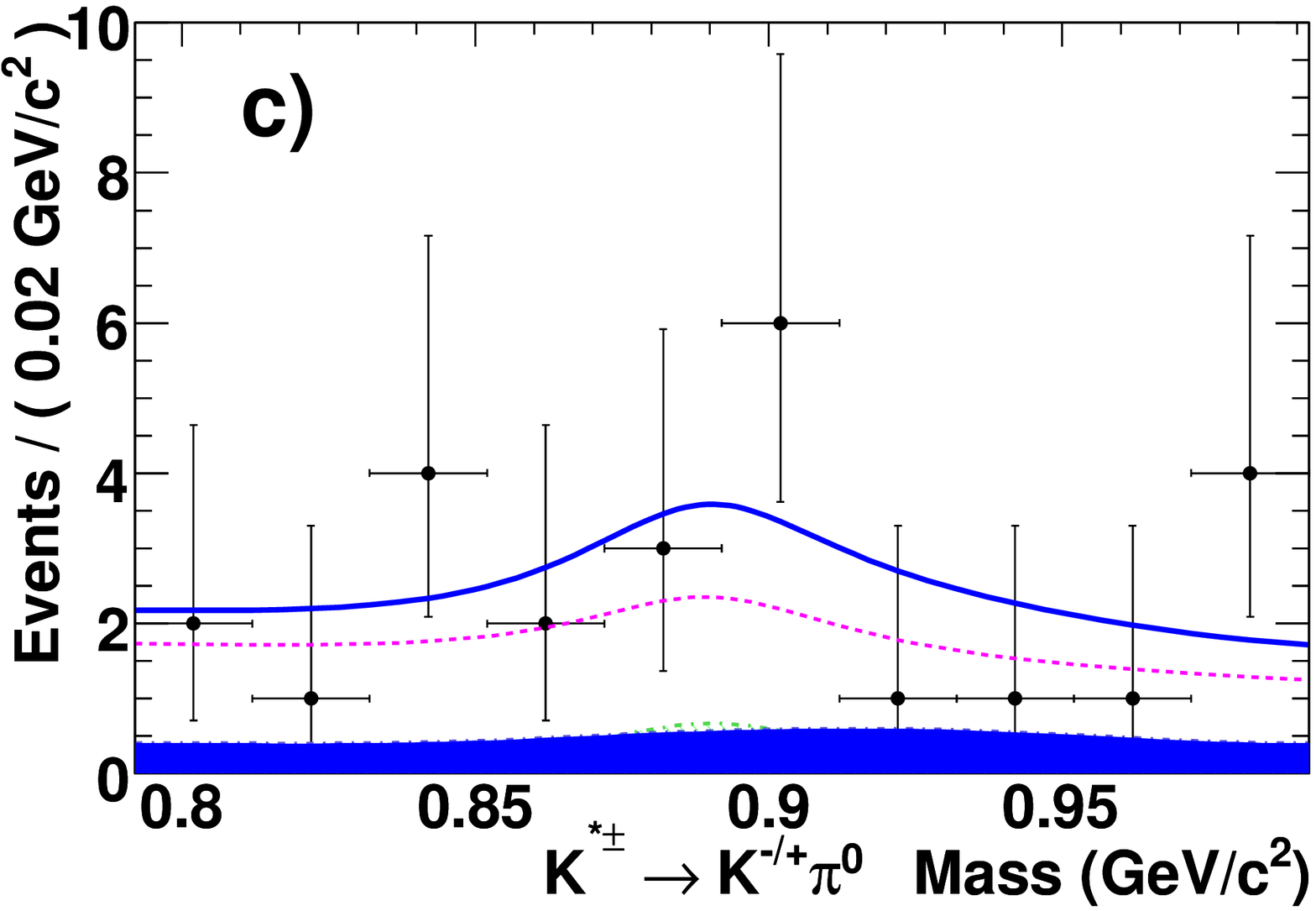}
\setlength{\epsfxsize}{0.5\linewidth}\leavevmode\epsfbox{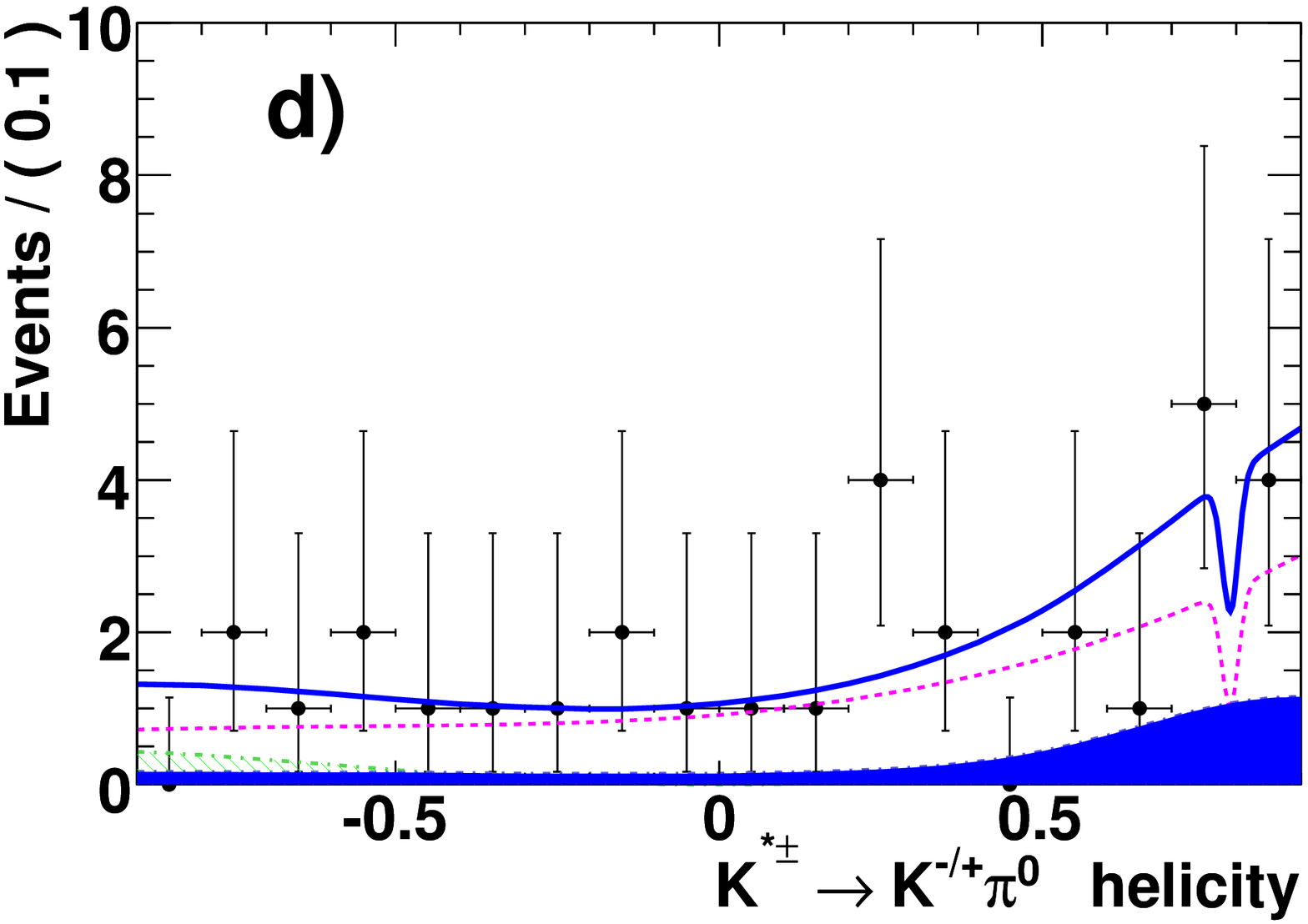}
}
\vspace{-0.3cm}
\caption{\label{fig:proj2} Projections of the
multidimensional fit onto (a) \mes; (b) \DeltaE; (c) \Kstarpm\ mass;
and (d) cosine of \Kstarpm\ helicity angle for \btoKstarpKstarmKp. The
same projection criteria and legend are used as in Fig.~\ref{fig:fig01}.}
\label{fig:fig02}
\end{figure}

Figures~\ref{fig:proj1} and~\ref{fig:proj2} show the projections of
the two fits onto \mes, \DeltaE, \Kstarpm\ mass and cosine of the
\Kstarpm\ helicity angle for the final state with zero and one \piz,
respectively.  The candidates in the figures are signal-enhanced with
a requirement on the probability ratio ${\cal P}_{\rm sig}/({\cal
P}_{\rm sig} +{\cal P}_{\rm bkg})$, optimized to enhance the
visibility of potential signal, where ${\cal P}_{\rm sig}$ and ${\cal
P}_{\rm bkg}$ are the signal and the total background probabilities,
respectively (computed without using the variable plotted).
The dip in helicity at $\sim 0.78$ is created by the criteria
 used to veto charm background.

\begin{table}[htb]
\caption{Summary of results with \fL=1.0 for the fitted yields, fit
  biases, reconstruction efficiencies $\epsilon$, sub-branching
  fractions $\prod\calB_{i}$, branching fraction
  \calB(\btoKstarpKstarm), significance $S$, and 90\% C.L. upper limit ${\cal B}_{\rm
  UL}$. The first error is statistical and the second, if given, is
  systematic.}
\begin{center}
\begin{tabular}{lcc}
\hline \hline
\noalign{\vskip1pt}
Final State  & \KS\pip\KS\pim & \KS\pipm\Kmp\piz \\ \hline
Yields (events):                   &   & \\
\; Total          & \kzntot & \kpntot\ \\
\; Signal         & \fkznsig & \fkpnsig \\
\; \BB\ bkg. & \fkznbb &  \fkpnbb \\
\; \qqbar\ bkg.     & \fkznuds & \fkpnuds \\
\; ML Fit Biases  & \kzbias  & \kpbias\ \\ \hline
Efficiencies and \calB: & & \\
\; $\epsilon(\%)$ & \kzeffl & \kpeffl  \\
\; $\prod\calB_{i} (\%)$ & 5.32 & 15.19 \\
%\; $\epsilon \times \prod\calB_{i} (\%)$  & \kzefflcor  & \kpefflcor \\
\; \calB\ ($\times 10^{-6}$) &\fkzbf &\fkpbf \\
%\; \fL\                   &    \kzfl &  \kpfl \\ 
\; Significance $S$ ($\sigma$) &  \fkzsig  &  \fkpsig \\ \hline
Combined Results: \\
\; \calB\ ($\times 10^{-6}$) & \multicolumn{2}{c}{\fkcbf} \\
\; Significance $S$ ($\sigma$)      & \multicolumn{2}{c}{\fkcsig} \\
\; ${\cal B}_{\rm UL}$ ($\times 10^{-6}$) & \multicolumn{2}{c}{\fkcup} \\
\hline
\hline
\end{tabular}
\label{tab:results}
\end{center}
\end{table}

%%%%%%%%%%%%%%%%%%%%%%%%%%%%%%%%%%%%%%%%%%%%%%%%%%%%%%%%%%%%%
% SYSTEMATIC STUDIES
%%%%%%%%%%%%%%%%%%%%%%%%%%%%%%%%%%%%%%%%%%%%%%%%%%%%%%%%%%%%%

The systematic uncertainties are summarized in Table~\ref{tab:syst}.
The errors on the branching fractions arise from the PDFs, fit biases
and efficiencies.  The PDF uncertainties are calculated by varying the
PDF parameters that are held fixed in the original fit by their
errors.  The uncertainty from the fit bias includes its statistical
uncertainty from the simulated experiments and half of the correction
itself, added in quadrature.  The uncertainties in PDF modeling and
fit bias are additive in nature and affect the significance of the
branching fraction results.  Multiplicative uncertainties include
reconstruction efficiency uncertainties from tracking and particle
identification (PID), track multiplicity, MC signal efficiency statistics,
and the number of \BB\ pairs.

\begin{table}[htb]
\caption{Estimated systematic errors in the final fit. 
Error sources which are correlated and uncorrelated 
when combined from the two decays 
are denoted by C and U, respectively.}
\begin{center}
\begin{tabular}{lccc}
\hline \hline 
\noalign{\vskip1pt}
Final State & \KS\pip\KS\pim & \KS\pip\Kpm\piz \\ \hline
Additive errors (events) & & \\
\; Fit Bias [U]            & 0.09 & 0.85 \\
\; Fit Parameters [U]      & 0.06 & 0.25 \\ \hline
Total Additive (events) & 0.10 & 0.88 \\ \hline
Multiplicative errors (\%)  & & \\
\; Track Multiplicity [C]  & 1.0  & 1.0 \\
\; MC Statistics  [U]      & 0.5  & 0.6 \\ 
\; Number of \BB\ pairs [C] & 1.1  & 1.1 \\
\; PID [C]                 & 2.2  & 1.1 \\
\; Neutrals Corrections [C]  & -  & 3.0 \\
\; \KS\ Corrections [C]      & 1.8  & 1.4 \\
\; Tracking Corrections [C]  & 1.6  & 0.8 \\  \hline
Total Multiplicative (\%) & 3.6 & 3.9 \\ \hline
Total \calB\ error ($\times 10^{-6}$) & 0.05 & 0.16 \\
\hline
\hline
\end{tabular}
\label{tab:syst}
\end{center}
\end{table}

%%%%%%%%%%%%%%%%%%%%%%%%%%%%%%%%%%%%%%%%%%%%%%%%%%%%%
% SUMMARY
%%%%%%%%%%%%%%%%%%%%%%%%%%%%%%%%%%%%%%%%%%%%%%%%%%%%%

In summary, we have measured the branching fraction ${\cal B}
(\btoKstarpKstarm) = [\fkcbf] \times 10^{-6}$, assuming the decay is
fully longitudinally polarized. The 90\% C.L. upper limit on the
branching fraction ${\cal B} (\btoKstarpKstarm) < \fkcup\times
10^{-6}$ is nearly two orders of magnitude more stringent than
previous searches.

%%%%%%%%%%%%%%%%%%%%%%%%%%%%%%%%%%%%%%%%%%%%%%%%%%%%%%%%%%%%
% ACKNOWLEDGMENTS
%%%%%%%%%%%%%%%%%%%%%%%%%%%%%%%%%%%%%%%%%%%%%%%%%%%%%%%%%%%%
%\input acknow_PRL.tex
\input acknowledgements.tex

%%%%%%%%%%%%%%%%%%%%%%%%%%%%%%%%%%%%%%%%%%%%%%%%%%%%%%%%%%%%
% BIBLIOGRAPHY
%%%%%%%%%%%%%%%%%%%%%%%%%%%%%%%%%%%%%%%%%%%%%%%%%%%%%%%%%%%%

\end{document}

%% file: numbers.tex
\newcommand{\onreslumi} {\mbox{413\invfb}}
\newcommand{\offreslumi} {\mbox{41.2\invfb}}
\newcommand{\nbb}       {\mbox{$454\pm5$}}

\newcommand{\kzntot}     {\mbox{$602$}}
\newcommand{\kznsig}     {\mbox{$1.8^{+2.7}_{-1.7} $}}

\newcommand{\kzfl}       {\mbox{$0.0 \pm 0.6 $}}

\newcommand{\kzbias}     {\mbox{$-0.170$}}
\newcommand{\fkznsig}     {\mbox{$0.7^{+2.5}_{-1.4} $}}
\newcommand{\fkznuds}     {\mbox{$580 \pm 25 $}}
\newcommand{\fkznbb}      {\mbox{$20 \pm 20 $}}
\newcommand{\fkzbf}       {\mbox{$0.38^{+1.1}_{-0.6} \pm 0.05 $}}

\newcommand{\fkzsig}      {\mbox{$0.50$}}

\newcommand{\kzeffl}     {\mbox{$8.89\pm0.08$}}

\newcommand{\kpntot}     {\mbox{$1923$}}
\newcommand{\kpnsig}     {\mbox{$4.1^{+5.8}_{-3.2} $}}

\newcommand{\kpfl}       {\mbox{$1.0 \pm 1.0$}}

\newcommand{\kpbias}     {\mbox{$1.70$}}
\newcommand{\fkpnsig}     {\mbox{$4.2^{+4.6}_{-3.3} $}}
\newcommand{\fkpnuds}     {\mbox{$1835 \pm 70 $}}
\newcommand{\fkpnbb}      {\mbox{$84 \pm 51 $}}
\newcommand{\fkpbf}       {\mbox{$0.76^{+1.4}_{-1.0} \pm 0.16$}}

\newcommand{\fkpsig}      {\mbox{$0.74$}}

\newcommand{\kpeffl}     {\mbox{$4.83\pm0.04$}}

\newcommand{\fkcbf}       {\mbox{$0.52^{+0.83+0.08}_{-0.58-0.06}$}}
\newcommand{\fkcup}       {\mbox{$2.0$}}
\newcommand{\fkcsig}      {\mbox{$0.87$}}

%% file: authors_may2008.tex
%% author list as of 05-May-2008 (530 authors)
%
\author{B.~Aubert}
\author{M.~Bona}
\author{Y.~Karyotakis}
\author{J.~P.~Lees}
\author{V.~Poireau}
\author{E.~Prencipe}
\author{X.~Prudent}
\author{V.~Tisserand}
\affiliation{Laboratoire de Physique des Particules, IN2P3/CNRS et Universit\'e de Savoie, F-74941 Annecy-Le-Vieux, France }
\author{J.~Garra~Tico}
\author{E.~Grauges}
\affiliation{Universitat de Barcelona, Facultat de Fisica, Departament ECM, E-08028 Barcelona, Spain }
% ...
\collaboration{The \babar\ Collaboration}
\noaffiliation

%% file: acknowledgements.tex
We are grateful for the 
extraordinary contributions of our \pep2\ colleagues in
achieving the excellent luminosity and machine conditions
that have made this work possible.
The success of this project also relies critically on the 
expertise and dedication of the computing organizations that 
support \babar.
The collaborating institutions wish to thank 
SLAC for its support and the kind hospitality extended to them. 
This work is supported by the
US Department of Energy
and National Science Foundation, the
Natural Sciences and Engineering Research Council (Canada),
the Commissariat \`a l'Energie Atomique and
Institut National de Physique Nucl\'eaire et de Physique des Particules
(France), the
Bundesministerium f\"ur Bildung und Forschung and
Deutsche Forschungsgemeinschaft
(Germany), the
Istituto Nazionale di Fisica Nucleare (Italy),
the Foundation for Fundamental Research on Matter (The Netherlands),
the Research Council of Norway, the
Ministry of Education and Science of the Russian Federation, 
Ministerio de Educaci\'on y Ciencia (Spain), and the
Science and Technology Facilities Council (United Kingdom).
Individuals have received support from 
the Marie-Curie IEF program (European Union) and
the A. P. Sloan Foundation.

%% file: paper.bbl
\begin{thebibliography}{99}

\bibitem{bib:ckm}
N.~Cabibbo, \jprl{10}, 531 (1963); 
M.~Kobayashi and T.~Maskawa, \progtp{49}, 652 (1973).

\bibitem{bib:Beneke06}
M.~Beneke, J.~Rohrer and D.~Yang, \npb{774}, 64 (2007).

\bibitem{cheng08}
H.Y.~Cheng and K.C.~Yang, arXiv:0805.0329v1.

\bibitem{bib:prediction}
A.~Ali {\em et al.}, Z.~Phys. C {\bf 1}, 269 (1979);
M.~Suzuki, \jprd{66}, 054018 (2002). 

\bibitem{bib:phiKst}
% phiK* BaBar/Belle
K.-F.~Chen {\em et al.} (Belle Collaboration), \jprl{94}, 221804 (2005);
B.~Aubert {\em et al.} (\babar\ Collaboration), \jprl{99}, 201802 (2007).

\bibitem{bib:KstKst}
B.~Aubert {\em et al.} (\babar\ Collaboration), \jprl{100}, 081801 (2008).

\bibitem{bib:theory1}
A.~Kagan, \plb{601}, 151 (2004);
C.~Bauer {\em et al.}, \jprd{70}, 054015 (2004);
P.~Colangelo {\em et al.}, \plb{597}, 291 (2004);
M.~Ladisa {\em et al.}, \jprd{70}, 114025 (2004);
H.-n.~Li and S.~Mishima, \jprd{71}, 054025 (2005);
M.~Beneke {\em et al.}, \jprl{96}, 141801 (2006).


\bibitem{bib:prevcleo}
R.~Godang \etal\ (CLEO Collaboration), \jprl{88}, 021802 (2001).

\bibitem{bib:kpkm}
 B.~Aubert \etal\ (\babar\ Collaboration), \jprd{75}, 012008 (2007);
 S.-W.~Lin \etal\ (Belle Collaboration), \jprl{98}, 181804 (2007);
A.~Bornheim \etal\ (CLEO Collaboration), \jprd{68}, 052002 (2003).

\bibitem{bib:babar}
B.~Aubert {\em et al.} (\babar\ Collaboration), \nima{479}, 1 (2002).

\bibitem{bib:polarization}
G.~Kramer and W.F.~Palmer, \jprd{45}, 193 (1992).

\bibitem{bib:PDG}
%Particle Data Group
W.-M.~Yao \etal\ (Particle Data Group), J. Phys. G {\bf 33}, 1 (2006).

\bibitem{bib:thrust}
S.~Brandt \etal, \jpl{12}, 57 (1964);
E.~Farhi, \jprl{39}, 1587 (1977).

\bibitem{bib:Legendre}
B.~Aubert {\em et al.} (\babar\ Collaboration),
\jprd{70}, 032006 (2004).

\bibitem{bib:tagging}
B.~Aubert {\em et al.} (\babar\ Collaboration),
\jprl{89}, 201802 (2002).

\bibitem{bib:argus} 
H.~Albrecht \etal\ (ARGUS Collaboration), \plb{241}, 278 (1990).

\bibitem{bib:geant}
S.~Agostinelli \etal\ (GEANT Collaboration), \nima{506}, 250 (2003).

\end{thebibliography}
